\documentclass[a4paper,12pt]{article}
\font\header=cmssdc10 at 20pt
\usepackage[ansinew]{inputenc}   
\usepackage[brazil]{babel}
\usepackage{amsfonts}
\usepackage{indentfirst}
  \usepackage[normalem]{ulem}
  \usepackage{graphicx}
  \usepackage{amssymb}
 \usepackage{draftcopy}
  \usepackage{amsmath}
  \usepackage{rotating}
  \usepackage{color}
  \usepackage{slashbox}
  \usepackage{setspace}
  \usepackage{fancybox}
 \usepackage{fancyvrb}
	\usepackage{fancyhdr}
	\usepackage[top=0.5cm,left=1cm,right=1cm,bottom=0.5cm]{geometry}
  \usepackage{wrapfig}

	\newcommand{\comenta}[1]{%
	} 

  \usepackage{float}
  \usepackage{type1cm}
\usepackage{eso-pic}
\usepackage{color}
 \usepackage{amsthm,amsmath,amsfonts}
\usepackage{fullpage,enumerate}
\usepackage{tikz}
\usepackage[colorlinks,citecolor=blue,urlcolor=blue]{hyperref}
\usepackage{hypernat}
\usepackage[notref,notcite]{showkeys}
\usepackage{graphicx}

\newcommand{\la}{\lambda}

\def\la{\lambda}

\begin{document}

{\header Survival under high mutation rates}

\vskip1cm

Rinaldo B. Schinazi

Department of Mathematics

University of Colorado at Colorado Springs

rschinaz@uccs.edu

\vskip1cm

{\bf Abstract} We consider a stochastic model for an evolving population. We show that in the presence of genotype extinctions the population dies out for a low mutation probability but may survive for a high mutation probability. This turns upside down the widely held belief that above a certain mutation threshold a population cannot survive. 

\vskip1cm

{\header 1 A model with genotype extinctions}

\vskip1cm

There seems to be a consensus in theoretical biology that mutations are helpful for the survival of a population but that too many mutations are not, see Nowak and May (2000), Eigen (2002) and Manrubia et al. (2010). We propose to use stochastic models to challenge this belief. In fact, we will show that there are situations where the population survives for large mutation probability but dies out for small mutation probability! That is, we propose to turn upside down the idea that too many mutations are necessarily bad. 

We now describe our model. Let $\mu$ be a continuous probability
distribution with support contained in $[0,\infty)$ and let $r\in[0,1]$.  Start with one individual at
time 0, and sample a birth rate $\la$ from the distribution
$\mu$. Individuals give birth at rate $\la$ and die at
rate 1. Every time there is a birth the new individual:
(i) with probability $1-r$ keeps the same birth rate $\la$ as its parent,
and (ii) with probability $r$ is given a new birth rate $\la'$, sampled
independently of everything else from the
distribution $\mu$. Furthermore, every time a new genotype appears (i.e. a new $\la$) we associate the genotype to a time $T$ (independently of everything else) sampled from a fixed distribution $\nu$. At time $T$ all the individuals of this genotype are killed and the genotype disappears from the population.

We think of $r$ as the mutation probability and
the birth rate of an individual as representing the genotype of the individual. 
Since $\mu$ is assumed to be continuous, a genotype cannot appear more than once in
the evolution of the population. Genotype extinctions happen when all the individuals with the same genotype die. These so called background extinctions have been going on since the beginning of life, see for instance Mayr (2001). 

We say that the population survives if there is a strictly positive probability of having at all times at least one individual alive. Note that no genotype can survive forever so the population may only survive if it generates infinitely many genotypes.

Our first result is a necessary and sufficient condition for survival of this population. We start the population with a single individual. 

\medskip

{\bf Theorem 1. }{\sl The population survives if and only if 
$$m(r)=rE[\Lambda\int_0^T\exp\big (\Lambda(1-r)-1)s\big )ds]>1,$$
where $\Lambda$ has distribution $\mu$ and $T$ has distribution  $\nu$.}

\medskip

We now use Theorem 1 to compute two limits. 

\medskip

{\bf Corollary 1. }{\sl Assume that
$$E[\Lambda\int_0^T\exp\big ((\Lambda-1)s\big )ds]<+\infty\leqno (1)$$
Then,
$$\lim_{r\to 0^+}m(r)=0.$$
$$\lim_{r\to 1^-}m(r)=E[\Lambda(1-e^{-T})].$$
}

\medskip

Note that hypothesis (1) holds true if for instance $\Lambda$ and $T$ have bounded support.

\medskip

{\it Proof of Corollary 1.}

For $r$ in $[0,1]$ we have
$$\exp\big ((\Lambda(1-r)-1)s\big )\leq \exp\big ((\Lambda-1)s\big).$$
Hence,  by the Dominated Convergence Theorem for fixed $\Lambda$ and $T$
$$\lim_{r\to 0^+}\int_0^T\exp\big ((\Lambda(1-r)-1)s\big )ds=\int_0^T\exp\big (\Lambda-1)s\big )ds$$
and
$$\lim_{r\to 1^-}\int_0^T\exp\big ((\Lambda(1-r)-1)s\big )ds=\int_0^T\exp(-s)ds=1-\exp(-T).$$

Observe now that 
$$\Lambda\int_0^T\exp\big ((\Lambda(1-r)-1)s\big )ds\leq \Lambda\int_0^T\exp\big ((\Lambda-1)s\big )ds.$$
By (1) the r.h.s. is integrable. Hence, the Dominated Convergence Theorem applies. We may interchange the limits in $r$ and the expectation. This yields the two limits and completes the proof of Corollary 1.

\medskip

We have the following consequences of Corollary 1. 

\medskip

$\bullet$  Under hypothesis (1) there exists $r_c$ in $(0,1)$ such that if $r<r_c$ then $m(r)<1$. Hence, by Theorem 1 survival is not possible for $r$ small. 

$\bullet$ If hypothesis (1) holds and we also have
$$E[\Lambda(1-e^{-T})]>1\leqno (2)$$
then there exists $r_c'$ in $(0,1)$ such that if $r>r_c'$ then $m(r)>1$ and survival is possible for $r>r'_c$.

\medskip

{\bf Remark 1.} Under hypotheses (1) and (2) survival is not possible for a low mutation probability but is possible for a high mutation probability. This goes exactly opposite to what is widely believed in Theoretical Biology. This belief has important practical consequences. One of the strategies to fight HIV has been to develop drugs to increase the mutation rate of the virus, see Manrubia et al. (2010). This can be futile or even counter productive if the virus can survive with increased mutation rate.

\medskip

{\bf Remark 2.} Hypotheses (1) and (2) can be met for a wide range of distributions. If for instance $\Lambda$ and $T$  are uniformly  distributed on $(0,a)$ and $(0,b)$, respectively, then (1) is true and (2) holds if and only if
$$\frac{a}{2}[1-\frac{1}{b}(1-e^{-b})]>1.$$
In particular, for any $a>2$ we can find $b$ large enough so that the inequality above holds.

\medskip

{\bf Remark 3.} It is not always true that the population dies out for small $r$. For instance,  if $\Lambda$ is the constant $\la$, $T$ is exponentially distributed with rate $\delta>0$ and $\la>1+\delta$ then $m(r)>1$ for all $r>0$. Hence, survival is possible for all $r>0$.

\vskip1cm

{\header 2 No genotype extinctions}

\vskip1cm

In this section we consider the model without genotype extinctions, everything else remains the same. That is, the population starts with a single individual, a birth rate is sampled from a fixed distribution $\mu$ and the death rate is 1. For every new individual the same birth rate is kept with probability $1-r$  or a new birth rate is sampled with probability $r$. With no genotype extinction a genotype can survive forever. In fact, if the birth rate is $\la$ for a particular genotype it will survive forever with positive probability if and only if $\la(1-r)>1$. This is so because the number of individuals with a fixed genotype is a birth and death process with birth rate $\la(1-r)$ and death rate 1. We have the following result.

\medskip

{\bf Theorem 2.} { \sl Consider a population with no genotype extinction. If the population has a positive probability of surviving for some probability mutation $r>0$ then there exists $r_c$ in $(0,1]$ such that the population survives for all $r<r_c$.}

\medskip

Theorem 2 shows that survival for high mutation and extinction for low mutation cannot happen without genotype extinctions.

\medskip

{\it Proof of Theorem 2}

Note first that if 
$$\mu\{\la:\la\leq 1\}=1$$
then the population dies out for all $r$ in $[0,1]$. This is so because we can couple our population to a birth and death chain with constant birth rate equal to 1 and death rate equal to 1. Since all the birth rates we sample for the population are below 1 it has less individuals than the birth and death chain with constant rates. Since the birth and death chain is critical it dies out and so does the population.

Assume now that our population has a positive probability of surviving for some $r>0$. Then we must have
$$\mu(\{\la:\la> 1\})>0$$
and therefore for some $s>0$
$$\mu(\{\la:\la(1-s)> 1\})>0.$$

Observe now that if a genotype has birth rate $\la$ such that $\la(1-s)>1$ then this genotype has a positive probability of surviving for any mutation probability $r\leq s$. This completes the proof of Theorem 2.

\vskip1cm

{\header 3 Proof of Theorem 1}

\vskip1cm

This proof is similar to the proof in Cox and Schinazi (2012). However, there the computation is done for the model with no genotype extinctions. Since the proof is short we include it for the sake of completeness. 
The main idea is the use of the genealogy tree of genotypes from Schinazi and Schweinsberg (2008). We now define this tree.

We say that the (unique) individual present at time zero has genotype $1$, and the $k$th type to appear will be called genotype $k$.
Each vertex in the tree will be labeled by a positive integer.  There will be a vertex labeled $k$ if and only if an individual of genotype $k$ is born at some time.  We draw a directed edge from $j$ to $k$ if the first individual of genotype $k$ to be born had an individual of genotype $j$ as its parent.  This construction gives a tree whose root is labeled $1$ because all genotypes are descended from  genotype $1$ that is present at time zero.  Since every genotype is eliminated eventually, the population  survives  if and only if the genealogy tree just described has infinitely many vertices.

We claim that this genealogy tree of genotypes is a discrete time Galton-Watson
tree. This is so because offsprings of different individuals in the tree are independent and have the same distribution. Let $m(r)$ be the mean offspring of a given genotype in this tree. We know this genealogy tree is infinite with positive probability if and only if $m(r)>1$, see Harris (1989). 

We now compute $m(r)$.
Recall that we start the population with a single individual with a genotype that we label 1. We associate a birth rate $\la$ and a death time $T$ to this genotype. Recall that $\la$ and $T$ are independent and sampled from fixed distributions $\mu$ and $\nu$, respectively.

Let $X_t$ be the number of genotype 1 individuals present at time $t$. Let $Y_t$ be the number of individuals born up to time $t$ that are offspring of genotype 1 individuals and underwent a mutation. We have for $t<T$
$${d\over dt}E(Y_t|\la)=\la rE(X_t|\la).$$
That is, genotype 1 individuals produce other genotypes at rate $\la r$. Since $X_t$ is a birth and death process with birth rate $\la(1-r)$ (genotype 1 individuals produce genotype 1 individuals at rate $\la (1-r)$)  and death rate 1 we have
$$E(X_t|\la)=\exp((\la(1-r)-1)t).$$
Hence, for $t<T$ we have
$$E(Y_t|\la)=r \la\int_0^t E(X_s|\la) ds=r\la\int_0^t \exp((\la(1-r)-1)s)ds .$$

On the other hand
$$E(Y_T)=\int E(Y_T|T=t)d\nu(t)=\int E(Y_t)d\nu(t).$$
Therefore,

\begin{align*}
E(Y_T)&=\int\int E(Y_t|\la)d\mu(\la)d\nu(t)\\
&=\int r \la\int_0^t E(X_s|\la) ds d\mu(\la)d\nu(t)\\
&=
\int r \la\int_0^t  \exp((\la(1-r)-1)s)ds d\mu(\la)d\nu(t).
\end{align*}

Hence,
$$E(Y_T)=rE[\Lambda\int_0^T\exp\big (\Lambda(1-r)-1)s\big )ds].$$

Note that $Y_T$ is the total number of genotypes that genotype 1 gave birth to before dying out. That is, it is the offspring of genotype 1 in the genealogy tree. Hence,
$$m(r)=E(Y_T).$$ 

This completes the proof of Theorem 1.

\vskip1cm

{\header References}

\vskip1cm

M. Eigen (2002) Error catastrophe and antiviral strategy. PNAS 99 13374-13376.

T.E.Harris (1989) {\it The Theory of Branching Processes.} Dover, New York.

S.C. Manrubia, E. Domingo and E. Lazaro (2010) Pathways to extinction: beyond the error threshold. Phil. Trans. R. Soc. B 365, 1943-1952.

E. Mayr (2001) {\it What evolution is.} Basic Books.

M. A. Nowak and R.M.May (2000) {\it Virus dynamics.}  Oxford University Press.

R.B. Schinazi (1999) {\it Classical and spatial stochastic processes.} Birkhauser.

R.B.Schinazi and J. Schweinsberg (2008) Spatial and non spatial stochastic models for immune response. Markov Processes and Related Fields 14 255-276.

\end{document}